\documentstyle[sprocl]{article}

\def\Journal#1#2#3#4{{#1} {\bf #2}, #3 (#4)}

\def\PLB{{\em Phys. Lett.}  B}
\def\PRL{\em Phys. Rev. Lett.}
\def\PRD{{\em Phys. Rev.} D}

\def\be{\begin{equation}}
\def\ee{\end{equation}}
\def\bea{\begin{eqnarray}}
\def\eea{\end{eqnarray}}
\def\bp{{\vec b}_\perp}
\def\qp{{\vec \Delta}_\perp}
\def\pp{{\vec p}_\perp}

\begin{document}

\title{OFF-FORWARD PARTON DISTRIBUTIONS AND 
IMPACT PARAMETER DEPENDENT PARTON STRUCTURE} 

\author{M. BURKARDT}

\address{Dept. of Physics, New Mexico State University, Las Cruces, NM 88003, 
USA,\\E-mail: burkardt@nmsu.edu}


\maketitle\abstracts{ 
The connection between parton distributions as a function 
of the impact parameter and off-forward parton 
distributions is discussed in the limit of vanishing skewedness parameter 
$\xi$, i.e. when the off-forwardness is purely transverse.
It is also illustrated how to relate
$\xi\neq 0$ data to $\xi=0$ data, which is important
for experimental measurements of these observables.}


\section{Introduction}
\label{introduction}
Deeply virtual Compton scattering experiments in the Bjorken
limit allow measuring generalized or
off-forward parton distributions (OFPDs) \cite{ji,ar}
\bea
\bar{p}^+\int \frac{dx^-}{2\pi} 
\langle p^\prime | \bar{\psi}(\frac{-x^-}{2}) \gamma^+
\psi(\frac{x^-}{2}) |p\rangle e^{ix\bar{p}^+x^-} 
&=&H(x,\xi,t)\bar{u}(p^\prime)\gamma^+ u(p)
\label{eq:off}\\
\nonumber
& &+E(x,\xi,t)\bar{u}(p^\prime)\frac{i\sigma^{+\nu}\Delta_\nu}{2M}u(p) 
, \\
\bar{p}^+\int \frac{dx^-}{2\pi} 
\langle p^\prime | \bar{\psi}(\frac{-x^-}{2}) \gamma^+\gamma_5
\psi(\frac{x^-}{2}) |p\rangle e^{ix\bar{p}^+x^-} 
&=&\tilde{H}(x,\xi,t)\bar{u}(p^\prime)\gamma^+\gamma_5 u(p)
\label{eq:offpol}\\
\nonumber
& &+\tilde{E}(x,\xi,t)\bar{u}(p^\prime)\frac{\gamma_5\Delta^+}{2M}u(p) 
, 
\eea
where $x^\pm = x^0\pm x^3$ and $p^+=p^0+p^3$ refer to the usual 
light-cone components, $\bar{p}=\frac{1}{2}\left(p+p^\prime\right)$,
$\Delta=p-p^\prime$, and $t\equiv \Delta^2$. The ``off-forwardness'' 
(or skewedness) in Eqs. (\ref{eq:off},\ref{eq:offpol}) is defined as 
$\xi\equiv \frac{\Delta^+}{p^+}$. From the point of view of parton 
physics in the infinite momentum frame, these OFPDs have the
physical meaning of the amplitude for the process that a quark 
is taken out of the nucleon with longitudinal momentum fraction $x$ 
and then inserted back into the nucleon with a four momentum 
transfer $\Delta^\mu$ \cite{wally}.
OFPDs play dual roles and in a certain sense they interpolate
between form factors and
conventional parton distribution functions (PDFs) \cite{ji,ar}:
for $\xi=t=0$ one recovers conventional PDFs, i.e.
longitudinal
momentum distributions in the infinite momentum frame (IMF), while
when one integrates $H(x,\xi,t)$ over $x$, one obtains a form factor,
i.e. the Fourier transform of a position space density 
(in the Breit frame!).
One of the new physics insights that one can learn from these
OFPDs is the angular momentum distribution \cite{hood}.
Others include meson distribution amplitudes. \footnote{For a discussion
of this connection in the context of $QCD_{1+1}$ see Ref. \cite{mb:2d}.}
Beyond that it is not yet entirely clear what kind of new physics 
insights one gains from studying these generalized 
parton distributions.
The main reason, why the physical interpretation of OFPDs is
still somewhat obscure is due to the fact that the initial and 
final state in Eq. (\ref{eq:off}) are not the same and therefore, in general, 
these OFPDs 
cannot be interpreted as a `density' but
rather their physical significance is that of a probability amplitude.

In this note, we will discuss the limit 
$\xi\rightarrow 0$, but $t\neq 0$, where 
\be
H(x,t)\equiv H(x,\xi=0,t)
\quad \quad \mbox{and} \quad \quad
\tilde{H}(x,t)\equiv \tilde{H}(x,\xi=0,t)
\ee
{\it do} have a simple interpretation in terms of a density, 
namely as the Fourier transform of impact parameter dependent
parton distributions w.r.t. the impact parameter, i.e.
\bea
H(x,-\qp^2)&=& \int d^2b_\perp q(x,\bp) e^{-i\bp \cdot \qp}
\nonumber\\
\tilde{H}(x,-\qp^2)&=& \int d^2b_\perp \Delta q(x,\bp) e^{-i\bp \cdot \qp}
\eea

\section{Impact parameter dependent PDFs
and OFPDs}
PDFs are usually defined as matrix elements
between plane wave states, which extend throughout space. 
Therefore, before we can introduce
the notion of impact parameter dependent PDFs, we need to
define localized wave packets. For our purposes, it is most
suitable to consider a wave packet $|\Psi \rangle$ which is 
chosen such that it has a sharp longitudinal momentum $p_z$, 
but whose position is a localized  wave packet in the 
transverse direction
\be
\left| \Psi \right\rangle = \int \frac{d^2p_\perp}
{\sqrt{2E_{\vec p}(2\pi)^2}} \Psi(\pp) \left|
{\vec p}\right\rangle .
\label{eq:Psi}
\ee
Although this definition is completely general, what we
have in mind for the states $|p\rangle$ are for example
nucleon states,
which are of course extended objects, i.e. Eq. (\ref{eq:Psi}) 
describes wave packets of particles that are themselves already
extended objects. We will come back to this point further below.
Clearly,
\be
F_\Psi(x,\bp) \equiv 
\int \frac{dx^-}{2\pi} 
\langle \Psi | \bar{\psi}(-\frac{x^-}{2},\bp) \gamma^+
\psi(\frac{x^-}{2},\bp) |\Psi\rangle e^{ix\bar{p}^+x^-} , 
\label{eq:fpsi}
\ee
describes the probability to find partons with momentum 
fraction $x$ at transverse (position) coordinate $\bp$ in this 
wave packet. Note that in Eq. (\ref{eq:fpsi}) we have implicitly
assumed that we work in light-cone gauge $A^+=0$. In any other 
gauge, one needs to insert a (straight line) `gauge string'
connecting the points $(-\frac{x^-}{2},\bp)$ and $(\frac{x^-}{2},\bp)$ 
in order to render Eq. (\ref{eq:fpsi}) manifestly gauge invariant.

What we will show in the following is that, for a suitably
localized wave packet, $F_\Psi (x,\bp)$ can be related to OFPDs
with $\xi =0$. Using Eq. (\ref{eq:Psi}), one finds
\bea
f_\Psi (x,\qp)&\equiv& \int d^2\Delta_\perp 
e^{-i\qp \cdot \bp} 
F_\Psi (x,\bp) \nonumber\\
&=& \int \frac{d^2p_\perp
\Psi^*(\pp^\prime) \Psi(\pp)}
{\sqrt{2E_{\vec p} 2 
E_{{\vec p}^\prime}}} 
\int dx^- e^{ix\bar{p}^+x^-} 
\langle p^\prime | 
\bar{\psi}(-\frac{x^-}{2},{\vec 0}_\perp) 
\psi (\frac{x^-}{2},{\vec 0}_\perp) |p\rangle
\nonumber\\
&=& \int \frac{d^2p_\perp
\Psi^*(\pp^\prime) \Psi(\pp)
}{\sqrt{2E_{\vec p} 2 
E_{{\vec p}^\prime}}}
H(x,\xi=0, t).
\label{eq:fourier}
\eea
where $\pp^\prime = \pp +
\qp$ and $p_z^\prime = p_z$, i.e. $\xi=0$. 
 
The physics of Eq. (\ref{eq:fourier}) is the following:
the non-trivial intrinsic structure of the target particle is
expressed in the OFPD $H(x,\xi,t)$. 
However, as the convolution of $H(x,\xi,t)$ with the wave
function reflects, the impact parameter dependent PD
in the state $\Psi$ gets spread out in position space
due to the fact that the wave function $\Psi$ does in general
have a nonzero width in position space.

Intuitively, one would like to chose a wave packet that is 
point-like in position space (i.e. constant in momentum space)
so that the $\bp$-dependence in Eq. (\ref{eq:fpsi}) is only
due to the intrinsic structure of the target particle but
not due to the wave packet used to `nail it down'.
However, one need to be careful in this step (and being able
to properly address these issues was the sole reason for
working with wave packets) because as soon as one localizes
a particle to a region of space smaller than its Compton
wavelength, its motion within the wave packet becomes
relativistic and therefore the structure of the particle
gets affected by Lorentz contraction as well as other
relativistic effects.

\subsection{Nonrelativistic limit}
It is very instructive to consider the nonrelativistic (NR)
limit first, where none of these complications occur.
Formally, the simplification arises since
$E_{\vec p}=E_{{\vec p}^\prime} =M$
and therefore $\Delta^2=-{\vec \Delta}^2 = -\qp^2$ 
First of all, this means that one can pull $H(x,0,-\qp^2)$
out of the integral in Eq. (\ref{eq:fourier}), yielding
\be
f_\Psi (x,\qp)= 
H(x,0, -\qp^2)\int \frac{d^2p_\perp
\Psi^*(\pp^\prime) \Psi(\pp)}{2M} .
\label{eq:fouriernr}
\ee
In order to proceed further, we choose a wave packet that is 
very localized in transverse position space. Specifically, we 
choose a packet whose width in transverse momentum space is 
much larger than a typical QCD scale. That way, the 
$\qp$-dependence on the r.h.s. of Eq. (\ref{eq:fouriernr}) is 
mostly due to the matrix element and not due to the wave 
packet $\Psi $. Therefore, by making the wave packet very 
localized in position space one obtains
\be
f_\Psi (x,\qp) =H(x, -\qp^2).
\ee
The dependence on the detailed shape of the wave packet has 
disappeared once it is chosen localized enough and it is thus 
legitimate to identify the Fourier transformed (w.r.t. $\qp$) 
$\xi=0$ OFPD with the impact parameter dependence of the 
parton distribution in the target particle itself.

\subsection{Relativistic corrections}
In order to have a unique definition of impact parameter 
dependent PDFs, i.e. a definition which does not depend on the 
exact shape of the wave packet,
one would like to make the wave packet as small as possible in position space
and certainly much smaller than the spatial extension of the hadron in the
$\perp$ direction --- otherwise Eq. (\ref{eq:fpsi}) is dominated by the
shape of the wave packet and not by the intrinsic $\bp$-dependence.
However, once the wave packet is smaller than about a Compton 
wavelength of the target then relativistic corrections 
can no longer be ignored and may even become more important as 
the corrections due to the spatial extension of the wave 
packet. 

This problem is well known from form factors, where the identification 
of the form factor with the Fourier transform of a charge distribution
in position space is uniquely possible only for momenta that are much
smaller than the target mass. One can also rephrase this statement in
position space by saying that the identification of the Fourier transform
of the form factor with a charge distribution works only if one looks at
scales larger than about one Compton wavelength of the target.
Details below this scale may depend on the Lorentz frame and are 
therefore not physical.

For OFPDs the problem is very similar. If one works for example in the rest
frame then one faces the same kind of corrections that also affect the form 
factors, in the sense that the $\perp$ resolution is limited to about
one Compton wavelength. More details can be found in Ref. \cite{me} and are 
omitted here since the natural frame to interpret parton distribution 
functions is the infinite momentum frame (IMF), and as we will 
discuss in the next section, these relativistic corrections 
play no important role in the IMF.

\subsection{Infinite momentum frame}
In the following we chose a wave packet as in Eq. (\ref{eq:Psi}) but
with $p_z\gg M$, where $M$ is the target mass. Then one can for 
example expand \footnote{Note that, by choice of the wave 
packet, ${\vec \Delta}\equiv {\vec p}^\prime -{\vec p}$ is always 
transverse i.e. $\Delta_z=0$.}
\bea
E_{{\vec p} + {\vec \Delta}} 
\approx |p_z| + \frac{\pp^2 + 2 \pp\qp + \qp^2}{2|p_z|}
\eea
and therefore the energy transfer
\be
E_{{\vec p} + {\vec \Delta}}-E_{{\vec p}} \sim 
\frac{2 \pp\qp + \qp^2}{2|p_z|}
\label{eq:deltaE}
\ee
vanishess as $p_z\rightarrow \infty$. A more detailed analysis 
works as follows: Let us denote the typical momentum scale in 
the wave packet by $\Lambda_{\pp}$
and the $\perp$ resolution that we are interested in by $\Lambda_{\qp}$ 
(normally, $\Lambda_{\qp}$ will be a typical
hadronic momentum scale, i.e. on the order of $1GeV$).
Then of course what we need to satisfy is $\Lambda_{\pp} \gg \Lambda_{\qp}$.
At the same time we need $|p_z| \gg \Lambda_{\pp}, \Lambda_{\qp}$, but this
is no problem if we go to the IMF.

As a direct consequence of Eq. (\ref{eq:deltaE}), the time-like
component of the momentum transfer can be ignored, i.e.
one can approximate $\Delta^2\approx-{\vec \Delta}^2 = -\qp^2$
and therefore one can again pull 
$H(x,0,-\qp^2)$ out of the integral in Eq. (\ref{eq:fourier}),
and one finds
\be
f_\Psi(x,\qp)= H(x,0,-\qp^2) \int \frac{d^2p_\perp
\Psi^*(\pp^\prime) \Psi(\pp)}{\sqrt{2E_{\vec p} 2 
E_{{\vec p}^\prime}}} \label{eq:imf}.
\ee
Finally making use of the fact that we chose a wave packet that is very
localized, i.e. $\Lambda_{\pp} \gg \Lambda_{\qp}$, we can 
approximate $\Psi(\pp^\prime) \approx \Psi(\pp)$ in Eq. 
(\ref{eq:imf}) yielding
\be
f_\Psi(x,\qp)= H(x,0,-\qp^2).
\ee
As in the NR case, the dependence on the wave packet 
has disappeared and the identification has become unique.

We have thus accomplished to show that, 
OFPDs at $\xi=0$ have in the IMF
the simple physical interpretation as Fourier transforms
of impact parameter dependent parton distributions with respect
to the impact parameter.
In other words, OFPDs, in the limit of $\xi\rightarrow 0$, 
allow to simultaneously determine
the longitudinal momentum fraction and transverse impact parameter
of partons in the target hadron in the IMF.

\section{The transverse center of momentum}
In the previous chapter, when we introduced the notion of 
PDFs as a function of the impact parameter, we
did not specify with respect to which point (or line) the impact 
parameter is defined. In NR scattering 
experiments, the impact parameter usually refers to the center 
of mass, but it is perhaps not a priori obvious how to 
generalize the concept of a $\perp$ center of mass to the IMF.

In the infinite momentum or light-front (LF) frame there exists 
a residual Galilei invariance under the purely kinematic $\perp$ boosts
\bea
x_i &\longrightarrow& x_i^\prime = x_i \nonumber\\
{\vec k}_{i\perp} &\longrightarrow& {\vec k}_{i\perp}^\prime
\equiv {\vec k}_{i\perp} + x_i \Delta {\vec P}_\perp,
\label{eq:imfboost}
\eea
where we denote the longitudinal momentum fraction and $\perp$ 
momentum of the $i$-th parton in a given Fock component by 
$x_i$ and ${\vec k}_{i\perp}$ respectively, the
LF-Hamiltonian transforms covariantly, i.e. \protect{
$P^-- \frac{{\vec P}_\perp^2} {2P^+}$} remains constant,
which resembles very much NR boosts
\be
{\vec k}_i \longrightarrow {\vec k}_i^\prime ={\vec k}_i
+ m_i \Delta {\vec v} =
{\vec k}_i+\frac{m_i}{M} \Delta {\vec P} ,
\label{eq:nrboost}
\ee
with $E-\frac{{\vec P}^2}{2M}$ remaining constant. Because of 
this similarity between NR boosts and $\perp$ boosts in the 
IMF, many familiar results about NR boosts can be immediately
transferred to the IMF. 

For this work, the most important result concerns the
physical interpretation of the form factor as the
Fourier transform of the charge distribution in a frame
where the center of mass
\be
{\vec R}_{CM} \equiv \sum_i \frac{m_i}{M} {\vec r}_i
\label{eq:rcm}
\ee
is at the origin. One can also rephrase this result by stating
that the Fourier transform of the form factor is the distribution
of the charge as a function of the distance from ${\vec R}_{CM}$.

By comparing Eqs. (\ref{eq:imfboost}) and (\ref{eq:nrboost})
it becomes clear that in the IMF the momentum fractions 
$x_i$ play the role of the mass fraction $\frac{m_i}{M}$.
Hence it is not very surprising that the IMF analog to the 
NR center of mass ${\vec R}_{CM}$ is given by the 
{\it transverse center of momentum} 
\be
{\vec R}_\perp \equiv \sum_i x_i {\vec r}_{i\perp} ,
\label{eq:rperp}
\ee
i.e. a weighted average of $\perp$ positions, but where the 
weight factors are given by the (light-cone) momentum fractions 
and not the mass fractions \footnote{Note that for NR systems 
the momentum fractions are given by the mass fractions.}.

The Fourier transform of the $\xi=0$ (i.e. `unskewed') OFPD
\be 
F(x,\bp)\equiv \int d^2\Delta_\perp
e^{i\bp \cdot \qp} H(x,0,-\qp^2)
\ee
can thus be interpreted as the (light-cone) momentum
distribution of partons as a function of the $\perp$
separation from the $\perp$ center of energy-momentum 
${\vec R}_\perp = \sum_i x_i {\vec r}_{i\perp}$ of the
target. \footnote{The above results can be verified by
starting with $\xi=0$ OFPDs calculated from a LF Fock 
space expansion for the hadron state \cite{kroll1}.}

The above observation about the $\perp$ center of momentum
has one immediate consequence for the $x\rightarrow 1$
behavior of $F(x,\bp)$. Since the weight factors
in the definition of ${\vec R}_\perp$ are the momentum 
fractions, any parton $i$ that carries a large fraction $x_i$
of the target's momentum will necessarily have a $\perp$ 
position ${\vec r}_{i\perp}$ that is close to ${\vec R}_\perp$.
Therefore the transverse profile (i.e. the dependence on
$\bp$) of $F(x,\bp)$ will necessarily
become more narrow as $x\rightarrow 1$, i.e. we expect that
partons become very localized in $\perp$ position as 
$x\rightarrow 1$. By Fourier transform, this also implies
that the slope of $H(x,0,t)$ w.r.t. $t$ at $t=0$,
i.e.
\be
\langle {\vec b}^2_\perp \rangle \equiv
4 \frac{\frac{d}{dt}\left. H(x,0,t)\right|_{t=0} }
{ H(x,0,0) }
\ee
should in fact vanish for $x\rightarrow 1$!

\section{Explicit example (model calculation)}
In order to illustrate the results from the previous sections
in a concrete example, we make a Gaussian ansatz for the
${\vec k}_{i\perp}$ dependence of the
light-cone wave function in each Fock component 
\cite{kroll1}
\be
\Psi(x_i,{\vec k}_\perp ) \propto
\exp \left[-a^2 \sum_{i=1}^N \frac{\left({\vec k}_{i\perp}
-x_i {\vec P}_\perp\right)^2}{x_i} \right] \Psi(x_i)
,
\label{eq:model}
\ee
where $\Psi(x_i)$ is not further specified. The above 
model ansatz allows to carry out the ${\vec k}_\perp$
momentum integrals explicitly, yielding
\be
H(x,0,t)=\exp\left[ \frac{a^2}{2} \frac{1-x}{x}t\right] f(x), 
\ee
i.e.
\be
F(x,\bp) \propto \frac{x}{1-x} 
\exp \left[-\frac{1}{2a^2}\frac{x}{1-x}\bp^2 \right] f(x).
\label{eq:final}
\ee
Of course, the above model (\ref{eq:model}) clearly 
oversimplifies the actual complexity of the nucleon's Fock
space wavefunction, but the qualitative behavior of the
final result (\ref{eq:final})
seems reasonable: as predicted in the previous section,
the width of $F(x,\bp)$ in $\perp$ position space
goes to zero as $x\rightarrow 1$. For decreasing $x$, the
distribution widens monotonically until it diverges for 
$x\rightarrow 0$. Of course, the divergence of the $\perp$
width as $x\rightarrow 0$ is probably an artifact of the model
ansatz which seems is more suitable for intermediate to large
values of $x$. Nevertheless, it seems reasonable that the 
$\perp$ width increases with decreasing $x$ since for example the 
pion cloud, which is expected to contribute only at smaller 
values of $x$, should me more spread out than the valence 
partons, which are expected to dominate at larger values of $x$. 

Notice also that both the above model calculation as well as the
pion cloud picture suggest that the $\bp$-dependence in $F(x,\bp)$
does  not factorize, which of course also translates back into a 
lack of factorization of the $t$ dependence in $H(x,0,t)$. A 
similar lack of factorization was also observed in Ref.
\cite{nonfac}.

\section{Practical aspects}
\label{sec:practical}
From the experimental point of view, $\xi=0$ is not directly accessible
in DVCS since one needs some longitudinal momentum transfer in order to 
convert a virtual photon into a real photon.
There are several ways around this difficulty. 
Of course, one way could be to access $\xi=0$ in real wide angle 
Compton scattering  \cite{ar3}. However, it should also be possible
to perform DVCS experiments at finite $\xi$ and to extrapolate
to $\xi=0$. 

Extrapolation to $\xi=0$ is greatly facilitated by working with moments
since the $\xi$ dependence of the
moments of OFPDs is in the form of polynomials \cite{wally}.
For example, for the even moments of $H(x,\xi,t)$ one finds\cite{hood}
\bea
H_{n}(\xi,t)&\equiv& \int_{-1}^1 \!\!\!dx x^{n-1} H(x,\xi,t)
\label{eq:moment}
= \sum_{i=0}^{\left[\frac{n-1}{2}\right]}\!\!
A_{n,2i}(t) \xi^{2i} + C_n(t) \\
&=& A_{n,0}(t) + A_{n,2}(t) \xi^2+...+ A_{n,n-2}(t) \xi^{n-2}
+C_n(t)\xi^n, \nonumber
\eea
i.e. for example
\be
\int_{-1}^1 \!\!\!dx x H(x,\xi,t)
=A_{2,0}(t) + C_2(t)\xi^2.
\ee
Since the $H_n$ have such a simple functional dependence on $\xi$,
one can thus use measurements of the moments of OFPDs at nonzero
values of $\xi$ to determine (fit) the ``form factors'' $A_{n,2i}(t)$ and 
$C(t)$. Once one has determined these invariant form factors, one can 
go back and evaluate Eq. (\ref{eq:moment}) for $\xi=0$, yielding
\be
H_n(\xi=0,t) = A_{n,0}(t),
\label{eq:a}
\ee
which then allows one to determine the impact parameter dependence
of the $n-th$ moment of the parton distribution in the target, using
\be
q_n(\bp) \equiv \int_{-1}^1 \!\!dx x^{n-1} q(x,\bp)
= \int d^2q_\perp A_{n,0}(-\qp^2)e^{i\qp \bp}.
\ee
A very similar procedure can be applied to the spin dependent OFPD
$\tilde{H}(x,\xi,t)$, whose even moments of $\tilde{H}(x,\xi,t)$ 
satisfy
\bea
\tilde{H}_{n}(\xi,t)&\equiv& \int_{-1}^1 \!\!\!dx x^{n-1} 
\tilde{H}(x,\xi,t)
\label{eq:moment2}
= \sum_{i=0}^{\left[\frac{n-1}{2}\right]}\!\!
\tilde{A}_{n,2i}(t) \xi^{2i} \\
&=& \tilde{A}_{n,0}(t) + \tilde{A}_{n,2}(t) \xi^2+...+ 
\tilde{A}_{n,n-1}(t) \xi^{n-1}. 
\nonumber
\eea
Similar to the unpolarized case, the $\frac{n+1}{2}$ form factors of 
the $n^{th}$ moment can be obtained from measurements of $\tilde{H}$ for $\frac{n+1}{2}$
different values of $\xi$ (and the same values of $t$)  
and the impact parameter dependence of the $n^{th}$ moment of the 
polarized parton distribution reads
\be
\Delta q_n(\bp) \equiv \int_{-1}^1 \!\!dx x^{n-1} \Delta q(x,\bp)
= \int d^2q_\perp \tilde{A}_{n,0}(-\qp^2)e^{i\qp \bp}.
\ee
Of course, this procedure becomes rather involved for high moments,
but the steps outlined above seem to be a viable way of determining
the impact parameter dependence of low moments of parton distributions
from DVCS data.

\section{Summary and outlook}
Off-forward parton distributions for $\xi\rightarrow 0$,
i.e. where the off-forwardness is only in the $\perp$
direction, can be identified with the Fourier transform of 
impact parameter dependent parton distributions w.r.t. the 
impact parameter $b$, i.e. the $\perp$ distance from the
center of (longitudinal) momentum in the IMF. This identification is very 
much analogous 
to the identification of the charge form factor with the
Fourier transform of a charge distribution in position space.

The $\xi\rightarrow 0$ limit of OFPDs is difficult 
to access in DVCS. However, as we illustrated  
in Sec.\ref{sec:practical}, one can also
measure $x$-moments for nonzero values of $\xi$ and use
those to construct moments for $\xi=0$. 

Knowing the impact parameter dependence allows one to gain information
on the spatial distribution of partons inside hadrons and to obtain
new insights about the nonperturbative intrinsic structure of hadrons.
For example, the pion cloud of the nucleon is expected to contribute more
for large values of $b$. Shadowing of small $x$ parton distributions,
is probably stronger at small values of $b$ since
partons in the geometric center of the nucleon are more effectively
shielded by the surrounding partons that partons far away from the 
center. These and many other models and intuitive pictures 
for the parton structure of hadrons give rise to predictions
for the impact parameter dependence of PDFs that reflect the
underlying microscopic dynamics of these models. Through the
experimental measurement of OFPDs using DVCS, it will for the
first time become possible to obtain experimental information
on the impact parameter dependence of parton structure 
which will thus provide much more
comprehensive tests on our understanding of nonperturbative
parton structure.

\section*{Acknowledgments}
This work was supported by a grant from
the DOE (DE-FG03-95ER40965) and in part by TJNAF.

\end{document}